\documentclass[journal=jacsat,manuscript=article,layout=twocolumn]{achemso}

\usepackage{geometry}
\usepackage[normalem]{ulem}

\setkeys{acs}{articletitle=true}
\usepackage{lettrine}

\usepackage[version=3]{mhchem} 
\usepackage[T1]{fontenc}       

\usepackage[fontsize=10pt]{scrextend}

\usepackage{newtxtext}
\usepackage{bm} 
\usepackage[labelfont=bf]{caption} 

\usepackage[hyperfootnotes=false]{hyperref}	 
\hypersetup{
  colorlinks,
  citecolor=blue,
  linkcolor=blue,
  urlcolor=blue
}

\usepackage{titlesec} 
\titleformat{\subsection}[runin]
{\normalfont\bfseries}{\thesubsection{.}}{1em}{}[.]

\titleformat{\section}
{\normalfont\sffamily\large\bfseries}{\thesection{.}}{1em}{\MakeUppercase}

\usepackage[table]{xcolor}
\usepackage{graphicx}
\usepackage{amsmath,fullpage,amssymb}
\usepackage{graphics,epsfig}
\usepackage{makeidx}
\usepackage{gensymb}
\usepackage{siunitx}

\newcommand{\kT}{k_\text{B} T}
\newcommand{\ksurf}{k_{{\text{r}}}}
\newcommand{\ktot}{k_{{\text{tot}}}}
\newcommand{\kapp}{k_{{\text{surf}}}}
\newcommand{\kd}{k_{{\text{D}}}}
\newcommand{\ep}{\varepsilon}
\newcommand{\lab}{\lambda_{{\text{AB}}}}
\newcommand{\rp}{r_{{\text{P}}}}

\usepackage{soul}	

\SectionNumbersOn
\let\oldmaketitle\maketitle
\let\maketitle\relax

\title{\flushleft Coverage Fluctuations and Correlations in Nanoparticle-Catalyzed Diffusion-Influenced Bimolecular Reactions}

\author{Yi-Chen Lin}
\affiliation{\rm\small Applied Theoretical Physics-Computational Physics, Physikalisches Institut, Albert-Ludwigs-Universit\"at Freiburg, Hermann-Herder Strasse 3, D-79104 Freiburg, Germany}
\email{yi-chen.lin@physik.uni-freiburg.de}

\author{Won Kyu Kim}
\affiliation{\rm\small Korea Institute for Advanced Study, 85 Hoegiro, Seoul 02455, Republic of Korea}
\email{wonkyukim@kias.re.kr}

\author{Joachim Dzubiella}
\affiliation{\rm\small Applied Theoretical Physics-Computational Physics, Physikalisches Institut, Albert-Ludwigs-Universit\"at Freiburg, Hermann-Herder Strasse 3, D-79104 Freiburg, Germany}
\alsoaffiliation{\rm\small Research Group for Simulations of Energy Materials, Helmholtz-Zentrum Berlin f\"ur Materialien und Energie, Hahn-Meitner-Platz 1, D-14109 Berlin, Germany}
\email{joachim.dzubiella@physik.uni-freiburg.de}

\begin{document}
\pagenumbering{arabic}
\noindent

\parindent=0cm
\setlength\arraycolsep{2pt}

\twocolumn[	
\begin{@twocolumnfalse}
\oldmaketitle

\begin{abstract}\small
The kinetic processes in nanoparticle-based catalysis are dominated by large fluctuations and spatiotemporal heterogeneities, in particular for diffusion-influenced reactions which are far from equilibrium.  Here, we report results from particle-resolved reaction-diffusion simulations of steady-state bimolecular reactions catalyzed on the surface of a single, perfectly spherical nanoparticle. We study various reactant adsorption and diffusion regimes, in particular considering the crowding effects of the reaction products. Our simulations reveal that fluctuations, significant coverage cross-correlations, transient self-poisoning, related domain formation, and excluded-volume effects on the nanoparticle surface lead to a complex kinetic behavior, sensitively tuned by the balance between adsorption affinity, mixed 2D and 3D diffusion, and chemical reaction propensity. The adsorbed products are found to influence the correlations and fluctuations, depending on overall reaction speed, thereby going beyond conventional steric (e.g., Langmuir-like) product inhibition mechanisms.  We summarize our findings in a state diagram depicting the nonlinear kinetic regimes by an apparent surface reaction order in dependence of the intrinsic reaction propensity and adsorption strength. Our study using a simple, perfectly spherical, and inert nanocatalyst demonstrates that spatiotemporal heterogeneities are intrinsic to the reaction-diffusion problem and not necessarily caused by any dynamical surface effects from the catalyst (e.g., dynamical surface reconstruction), as often argued. 
\vspace{5ex}
\end{abstract}
\end{@twocolumnfalse}]

\maketitle
\setlength\arraycolsep{2pt}
\small

\section{Introduction}

Nanoparticle-based catalysis in the liquid phase is a rapidly growing field with a wide spectrum of applications, ranging from environmental remediation to the selective production of sustainable fuels, medical drugs, and many other functional chemicals.\cite{bell:science, jin:nature, astruc, green, herves2012}  Due to the small size of the well-dispersed nanoparticles, the field of nanoparticle catalysis is considered somewhere between homogeneous and heterogeneous catalysis.~\cite{astruc} Indeed many of the kinetic phenomena known for surface reactions in traditional heterogeneous catalysis~\cite{ertl:book} qualitatively change if reactions are limited to the small sizes and low numbers of reactants at the nanoscale, e.g., as demonstrated early for surface supported nanocatalysts.~\cite{Zhdanov2000} In addition to the increasingly important effects of facets and defects on the highly heterogeneous nano-surfaces, in particular the role of spatiotemporal nano-fluctuations, scaling $1/N^{1/2}$ with system size,  becomes increasingly significant.~\cite{Kang1985, Zhdanov2002, imbihl2003, chemical, hinrichsen} 

In that respect it was indeed shown experimentally~\cite{nanofluct,imbihl2003, Libuda, grosfils} that reactant coverage fluctuations on supported model catalysts can drastically alter their macroscopic catalytic behavior. It was demonstrated  that macroscopically observable bistabilities, i.e., the existence of two stable kinetic regimes co-existing for a given set of reaction conditions,\cite{Ziff1986, Zhdanov1994}  vanish completely with decreasing particle size. The effect was attributed to fluctuation-induced transitions between the kinetic reaction regimes, controlled by both particle size and surface defects.~\cite{nanofluct,imbihl2003, Libuda, grosfils} These results suggest that fluctuation-induced effects represent a general phenomenon affecting the reaction kinetics and dynamical structure in nano-sized systems. Major recent advances in single-nanoparticle chemical imaging methods,\cite{Buurmans} for example, fluorescence microscopy~\cite{Peng:Nature:2008,Peng:PCCP:2009,Peng:JACS:2010} revealed indeed large spatiotemporal heterogeneities and activity fluctuations on individual metal nanoparticles, apparently attributable to both catalysis-induced and spontaneous dynamic surface restructuring that occur at different timescales at the catalytic surface. 

In general, irreversible dynamical reaction systems are known to show highly intricate behavior, including dissipative structures, domain formation,  oscillations, different adsorption states, kinetic phase transitions, etc.,~\cite{ertl:book, Ziff1986, Zhdanov1994,cross1993,wintterlin1997, shvartman1999, wintterlin2000, ertl1991, Zhdanov2002} a prominent and well studied example being the CO-oxidation reaction.~\cite{ertl:book,co:jacs,co:nature}  Experimental identification and quantification of noise-induced effects both in the global and micro-reaction kinetics of a nano-sized system is thus a rather difficult problem. Historically, mostly stochastic computer simulations~\cite{Andy, Ziff1986, Fichthorn, jensen, blumen, Zhdanov1994, Zhdanov2002, erban, Readdy, andrews2004, Liu:review, stamatakis} and theory, e.g., typically mean-field reaction-diffusion equations or master-equation approaches,~\cite{kuzovkov1988, Zhdanov1994, Zhdanov2002, freeman,chemical, hildebrand1996, gopich, hildebrand2003, pineda, grosfils,Bolhuis2,Bolhuis,roa2017} have contributed to our molecular-level understanding of fluctuation and correlation effects in chemical reaction kinetics.

Our simulation study is motivated by a few urgent, still-open questions related to diffusion-influenced nanoparticle-catalysis in the liquid phase.  
First of all, previous experimental works attribute dynamical heterogeneities often to the dynamical changes of the nanoparticle surface (such as reconstruction) but do not differentiate this from the intrinsic reaction-diffusion heterogeneities, in particular for diffusion-influenced reactions,~\cite{smoluchowski1906, Debye1942,Calef1983} which are far from equilibrium.  Also, previous theoretical studies on nano-sized, finite systems, dealt only with planar two-dimensional (2D) surfaces with an underlying lattice structure, often communicating with a surface support.~\cite{Zhdanov2000} Typically, only gas-phase reactions were considered assumed to occur with direct random adsorption from a vapor reservoir, but no desorption kinetics on the 2D surface was considered. Diffusive surface relaxation was included only in certain extreme limits (infinite relaxation or immobile) with respect to the intrinsic reaction scale.~\cite{blumen, Zhdanov2002} Apart from notable exceptions using approximative diffusion theory for ideal reactants (and no products) on planar surfaces,\cite{freeman} the effects of coupled 3D/2D diffusive search and adsorption/desorption on the total reaction rate were thus hardly addressed in literature. In particular, the effects of surface curvature of a spherical catalyst in 3D were not considered in theory nor in particle-resolved simulation studies up to date. Moreover, previous studies assumed instantaneous desorption (and the vanishing) of all products after the reaction,~\cite{Fichthorn, jensen} and the action of reversibly desorbing/adsorbing products and related crowding effects\cite{Roa2018} have not been studied so far for bimolecular surface reactions. Hence, the intrinsic roles of fluctuations and correlations as well as products in bimolecular reactions catalyzed by a single nanoparticle are still not well understood.

Typical nanoparticle catalyzed reduction and oxidation reactions are bimolecular,\cite{Atkins} that is, two reaction partners by processes of 3D diffusion, adsorption, and 2D diffusion have to find each other on the catalytic surface for their chemical transformation to occur and products to form.  Hence, we consider a general type of nanoparticle catalyzed bimolecular reactions, where the adsorption, desorption, and the irreversible reaction of reactants (A$\equiv$B) on the nanoparticle catalyst surface can be described by the Langmuir-Hinshelwood mechanism~\cite{Langmuir1916,Hinshelwood, Atkins}
\begin{equation}
\begin{split}
\text{A}+\text{catalyst} &\overset{k_{\rm a}}{\underset{k_{\rm d}}\rightleftharpoons} \text{A}_{{\text{ad}}},\\
\text{B}+\text{catalyst} &\overset{k_{\rm a'}}{\underset{k_{\rm d'}}\rightleftharpoons}  \text{B}_{{\text{ad}}},\\
\text{A}_{{\text{ad}}} + \text{B}_{{\text{ad}}} &\overset{k_{{\text{surf}}}}{\longrightarrow} \text{C}_{{\text{ad}}} + \text{D}_{{\text{ad}}} \;\; (\text{products}),\\
\end{split}
\label{eq:chem}
\end{equation}
which incorporates adsorption (with rates $k_{\rm a}$ and $k_{\rm a'}$), desorption (with rates $k_{\rm d}$ and $k_{\rm d'}$) and in general a 3D/2D diffusive search on the spherical nanoparticle for the  reaction transformations  between the interacting particles arranging on the surface.  
Diffusive and related local search processes are absorbed in the rate constants, in particular $k_{\rm surf}$ carries information on 2D diffusive search and crowding/interactions on the surface.  These highly correlated reaction processes are clearly  beyond elementary reaction kinetics and the reaction order (with respect to reactant bulk concentrations) can typically not be defined.~\cite{wunder2011}  The exact rate equation and more microscopic dissections of $k_{\rm surf}$ are currently not known, to the best of our knowledge. Hence, it is assumed almost exclusively in theoretical approaches for heterogeneous surface reactions that the total rate of the bimolecular reaction is given by a mean-field relation simply linear in the product of individual surface coverages~\cite{ertl:book, Atkins, Zhdanov1994,Zhdanov2002,wunder2011, roa2017}
\begin{equation}
\label{eq:ratelin}
 {\rm d}n/{\rm d}t = \kapp \theta_{{\text{A}}} \theta_{{\text{B}}},
\end{equation}
where $n$ is the number of reaction events,  $\kapp$ is the apparent rate constant for surface reactions, and $\theta_\text{A}$ and $\theta_\text{B}$ represent the surface coverages of the adsorbed reactants.  In other words, it is commonly assumed that the reaction is first order with respect to the individual surface coverages of the adsorbed reactants. This assumption is the basis for the standard Langmuir-Hinshelwood rate equations in the reaction-controlled limit (where diffusion limitation plays no role), and the coverages (adsorptions) can be conveniently expressed by equilibrium Langmuir-isotherms.~\cite{Atkins,enzyme}  Note that ansatz (2) is more general, and it makes no equilibrium assumption. However, for  strongly correlated and fluctuating systems on nano-scales eq.~(2) is expected to break down in certain regimes, where many-body collective correlation effects, i.e., cross-correlations between the coverages, may come into play.\cite{kuzovkov1988}

Here, using (off-lattice) particle-resolved reaction-diffusion (PRRD) computer simulations, we study the influence of fluctuations and correlations on bimolecular reactions catalyzed by a model nanoparticle, with a particular focus on the effects of adsorption strength, surface diffusion, and the presence of products. In our model the nanoparticle catalyst is a perfect sphere to probe only the intrinsic reaction-diffusion effects independent from any surface-mediated effects (e.g., dynamical surface reconstruction). The reactant coverage is then the decisive parameter, which in our work is tuned by adsorption energies of the reactants to the catalyst surface (acting like an inverse temperature).   We simulate the model system in a steady-state and calculate the mean transformation rate.  We find a rich scenario of nonlinear regimes, where the relation eq.~(2) is violated for different physical reasons. In particular, we observe the highly nonlinear behavior for adsorption energies $\gtrsim 5~ \kT$, i.e., large coverages,  either due to packing correlations for slow reactions, amplified by products, or by spatial domain formation and instabilities for very fast reactions, also significantly modified by the presence of products. In the presence of products we also observe the phenomenon of bifurcation, i.e., two different total rates are observed for the same product of surface coverages $\theta_\text{A} \theta_\text{B}$.  We finally summarize our findings in a state diagram depicting the nonlinear kinetic regimes in dependence of the intrinsic reaction rate~\cite{Bolhuis2, Hoefling} and adsorption strengths. 

\section{Model and simulation methods}

\subsection{Model for nanoparticle-catalyzed bimolecular reactions}

\begin{figure*}[h!]
 \includegraphics[width = 1.0\textwidth]{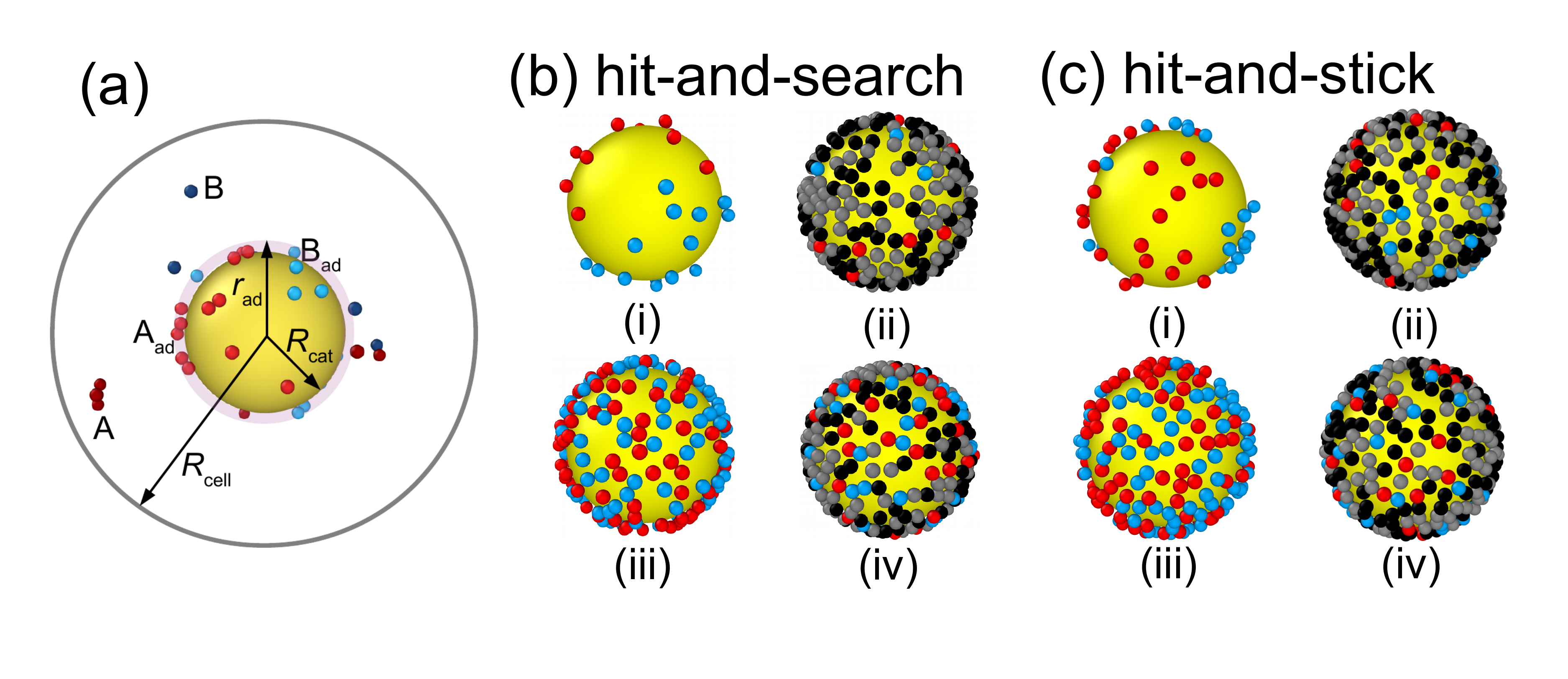}
\caption{Simulation model and representative snapshots. (a) Simulation cell for bimolecular reactions catalyzed at the surface of a nanoparticle catalyst (central yellow sphere): particles A (dark red) and B (dark blue) diffuse in the 3D spherical cell of radius $R_{\rm cell}=5.5$~nm. They can adsorb on the catalyst of radius $R_{\rm cat}=2$~nm with energy $\varepsilon$ to form the reactive reactants $\text{A}_\text{ad}$ (light red)  and $\text{B}_\text{ad}$ (light blue). Two reactants can react only if adsorbed on the catalyst surface and located within a mutual distance $r_{\rm reac}$ with a given probability defined by $\ksurf$ (see text and Methods) to either vanish or form product particles (not shown).  On the catalyst surface the reactants can diffuse with bulk diffusivity ('hit-and-search'), or, in a second scenario, are immobile after adsorption ('hit-and-stick') but can reversibly desorb.  (b) Simulation snapshots of reactants (red and blue) and products (gray and black) on the catalytic sphere for different $\ksurf$ and $\ep$ in the hit-and-search scenario.   (i) $\ksurf = \infty, \ep = 10\ \kT$, without products. (ii) $\ksurf = \infty, \ep = 10\ \kT$, with products. (iii) $\ksurf = 1\ \kd, \ep = 10\ \kT$, without products.  (iv) $\ksurf = 1\ \kd, \ep = 10\ \kT$, with products. (c) Same as (b) but now for the hit-and-stick case. For better visibility the particles in the bulk are not shown in the snapshots.}
\label{fig:system}
\end{figure*}

Our simulation model is illustrated in Fig.~\ref{fig:system}(a).  We consider a solid perfect sphere representing a nanoparticle catalyst of radius $R_{\rm cat}$ = 2 nm at the center of a spherical simulation cell of radius $R_{\rm cell} = 5.5$~nm. Two reactant types $\text{A}$ and $\text{B}$  with total number $N_{\rm A} = N_{\rm B} = 250$ are simulated. The initial average density of free A (or B) is thus $\rho_{\rm A(B)}^{\rm 0} = 0.3768$ nm\textsuperscript{-3}. A Lennard-Jones (LJ) potential $U_{\rm LJ}(r_{ij}) = 4\varepsilon_{ij}[(\sigma/r_{ij})^{12}-(\sigma/r_{ij})^6]$ is used for interactions between the reactants $i$ and $j$, where $r_{ij}$ is the distance between particle $i$ and $j$, $\ep_{ij} = 0.1~\kT$ is the  LJ interaction energy   and $\sigma = 0.35$~nm is the diameter of every species.  The interaction between the catalytic sphere and the reactants is considered by a displaced LJ potential $U_{\rm LJ}^{\rm S}(r_i) = U_{\rm LJ}(r_i - R_{\rm cat})$, where $r_i$ is the distance of particle $i$ to the catalyst center,  with an interaction energy $\ep_{\rm cat, A} = \ep_{\rm cat, B} \equiv \ep$, i.e., we consider A and B to have the same interaction properties.  The energy $\varepsilon$ essentially acts as an inverse temperature for the adsorption/desorption equilibrium and we use it as a free parameter to tune the adsorption and the resulting surface coverages.

The reactants A and B diffuse freely within the spherical cell both with the same diffusion constant $D$ and can reversibly adsorb (and desorb) on the catalyst surface subject to the displaced LJ interaction potential $U_{\rm LJ}^{\rm S}(r_i)$. We consider adsorption (desorption) occurs if the particle distance $r_i$ is smaller (larger) than the adsorption cut-off radius $r_{\rm ad} = R_{{\text{cat}}} + 1.2445\sigma$, which presents the location of the inflection point of $U_{\rm LJ}^{\rm S}(r_i)$. The dimensionless coverages (that is, area packing fractions) in our model are then defined by \begin{eqnarray}
\theta_i = N_{i,\rm {ad}}\pi\sigma^2/(16\pi R_{\rm cat, eff}^2),
\end{eqnarray} 
that is, by the ratio of the area of the $N_{i,\rm {ad}}$ adsorbed particles to the total available area on the catalyst surface with effective radius $R_{\rm cat, eff} = R_{\rm cat} + 2^{1/6}\sigma$. In the adsorbed state a particle becomes reactive and we assign A$\rightarrow$ A$_{\rm ad}$ (or B$\rightarrow$ B$_{\rm ad}$). Only in the adsorbed states they can react according to scheme (1) with details described in the next subsection. 

We study two scenarios, one in which the adsorbed particles can diffuse on the nanoparticle surface with unmodified bulk diffusivity $D$ (until they desorb or react), that is, a {\bf 'hit-and-search'} scenario, and a second scenario where the particles will be immobilized at the position where they land, i.e., a {\bf 'hit-and-stick'} scenario. Note that in the latter they can still desorb, though, so that a 3D 'hopping' search is still feasible. For a real nanoparticle in experiments these two scenarios would be generated by extremely low or high diffusion barriers for surface diffusion, respectively.  

In order to characterize the action of the interacting products we consider two cases: First, as shown in the representative snapshots in Figs.~\ref{fig:system}(b) (i), (iii) and (c) (i),(iii),  we consider the bimolecular reaction {\bf without products:} The reactants $\text{A}_\text{ad}$ and $\text{B}_\text{ad}$ are immediately removed from the nanoparticle surface after the reaction,  and randomly re-inserted at the inner boundary of $r_i = R_{{\text{cell}}}$, thereby ensuring a steady-state in the simulation. This simulates the limiting situation where the products readily desorb from the surface right after the transformation and quickly (here, actually instantaneously) diffuse away.   We compare this to the second case {\bf with products}, as shown exemplarily in Figs.~\ref{fig:system}(b) (ii),(iv) and (c) (ii),(iv). Here, the reactants $\text{A}_\text{ad}$ and $\text{B}_\text{ad}$ are transformed into non-reactive products $\text{C}_\text{ad}$ and $\text{D}_\text{ad}$ during the reaction, respectively.   For simplicity, we assume they have identical interactions as the reactants, in particular the same attraction $\varepsilon$ to the catalyst surface, i.e., they can desorb/adsorb reversibly  like the reactants.  To maintain a steady-state in the simulation with the products, the latter are transformed back to reactants A and B when they reach the boundary of the simulation cell.

\subsection{Reaction-diffusion simulations}

For the numerical solution of our model we perform (overdamped Langevin) Brownian dynamics (BD) simulations including molecular reactions~\cite{Gillespie, erban, Readdy}. Thus, the reactants are modeled as explicitly resolved, diffusing solutes in a viscous continuum background (the solvent).
The position of $j$-th particle $\bm{X}_{j}$ at time $t$ is computed by numerically iterating the overdamped Langevin equation using a Euler-Maruyama scheme~\cite{Ermak}
\begin{equation}
\gamma \frac{ \partial \bm{X}_{j}(t) }{ \partial t }=  -\nabla_j U_\text{tot} + {\bf R}_j(t), 
\end{equation}
where $\gamma$ is the friction constant, $U_\text{tot} = \sum_{i} \left[ U_\text{LJ}(\bm{X}_{i}) + U_\text{LJ}^\text{S}(\bm{X}_{i}) \right]$ 
is the total potential, and ${\bf R}_j(t) = (R_j^x, R_j^y, R_j^z)$ is the Gaussian random force, which satisfies the fluctuation-dissipation relation 
$\langle R_j^\alpha (t) R_j^\beta (t')\rangle = 2 \gamma \kT \delta_{\alpha \beta}\delta(t - t')$ and has the zero mean $\langle R_j^\alpha (t) \rangle = 0$. In the hit-and-stick case and additional, position-restraining angular (i.e., acting only parallel to the NP surface) harmonic potential with a stiff spring constant $k=1000~\kT/{\rm nm^2}$ is applied to the nanoparticle-adsorbed reactants to immobilize them. In the simulations we set units for the energy to $\kT$, the length to nm, and the time to the Brownian time scale $\tau_{\rm B}$, such that the diffusion constant is $D=\kT/\gamma = 1$~nm$^2/\tau_{\rm B}$.  The simulation time step $\Delta t = 10^{-5}\tau_{\rm B}$ is used, and the simulations are performed up to $10^3\tau_{\rm B}$. As a cutoff length of the LJ potential $2.5\sigma = 0.875$~nm is used. 

Reaction events can only occur on surface-adsorbed reactants and are handled according to the Doi model:~\cite{Doi:1975,andrews2004,erban,Hoefling}  When the distance between $\text{A}_\text{ad}$ and $\text{B}_\text{ad}$ on the catalytic surface is within a distance $r_\text{reac}=\sqrt[6]{2}\sigma$,  an irreversible reaction occurs with the intrinsic reaction rate constant (propensity\cite{Gillespie,Doi:1975,erban,Bolhuis2,Hoefling}) $\ksurf = P / \Delta t$, where $P = \ksurf\Delta t$ is a reaction probability  in the time step $\Delta t$ for  A$_{{\text{ad}}}$ and B$_{{\text{ad}}}$ to react and vanish (in the systems without products) or turn into products C$_{{\text{ad}}}$ and D$_{{\text{ad}}}$ (in the systems with products). The intrinsic rate $k_{\rm r}$ will be another free parameter which we use to interpolate between diffusion-controlled and more reaction-controlled regimes \cite{Calef1983,CK,Hoefling}. 

In our study we investigate intrinsic rate constants $\ksurf=1$, 10, 100, and $\infty$ in units of the Smoluchowski diffusion rate $\kd$.  The latter is the mean rate for a single particle to reach the nanoparticle surface and we define as
\begin{equation}
 k_{\text{D}} = \frac{1}{\tau_{\text{D}}(\varepsilon)},
\end{equation}
where $\tau_{\text{D}}(\varepsilon)$ is the mean-first passage time (MFPT) of a single particle diffusing from $r = R_{\text{cell}}$ to $r = r_{\text{ad}}$ with a reflecting boundary condition applied at $r = R_{\text{cell}}$ \cite{Szabo}
\begin{equation}
 \tau_{\text{D}}(\varepsilon) = \int_{r_{\text{ad}}}^{R_{\text{cell}}}\text{d}r''\frac{\text{exp}\left[\beta U_{\text{LJ}}(r'', \varepsilon)\right]}{Dr''^2}\int_{r''}^{R_{\text{cell}}}\text{d}r'r'^2\text{exp}\left[-\beta U_{\text{LJ}}(r', \varepsilon)\right], 
\end{equation}
with $\varepsilon$ is chosen to be the largest considered value of 10 $\kT$ to estimate the fastest diffusion-controlled limit for the total rate in our simulations. For infinite large cells, $\kd$ is defined by the classical Debye-Smoluchowski limit for unimolecular diffusion-controlled reactions.~\cite{smoluchowski1906,Debye1942,Calef1983}. For the size of our simulation cell we find $\tau_{\rm D} = 7.302~\tau_{\rm B}$ for $U_{\rm LJ}=0$ which is the MFPT for reaching the nanoparticle surface at $r_{\rm ad}$.  Hence, $\kd$ is on the order of $\simeq 0.14\tau_{\rm B}^{-1}$ and therefore $P = \ksurf\Delta t$ is in the range of 10$^{-5}$ to 10$^{-3}$. This is well below unity and the reaction probability for discrete simulations $P = 1 -\exp(-\ksurf\Delta t)$ (Poisson probability of finding at least one reaction event with rate $\ksurf$ in a time window $\Delta t$)\cite{vankampen,Readdy} well approximated with the linear relation $P = \ksurf\Delta t$. To simulate the limit $\ksurf =\infty$, we simply impose $P=1$.~\cite{andrews2004}

We would like to note that the propensity $k_{{\text{r}}}$ must not be confused with the overall surface reaction constant $k_{\rm surf}$, as in eq.~(2), which includes in addition to the chemical propensity the effects of diffusive surface search. The relation between $k_{\rm r}$ and $k_{\rm surf}$ is complex and not known for our model. Analytical solutions exist for the Doi model for 3D bulk bimolecular reactions~\cite{erban} with recent extension to interacting systems.~\cite{Hoefling} Approximate solutions exist for 2D planar surface reaction coupled to incoming perpendicular flux using diffusion equations.~\cite{freeman} In mean-field approaches to nanoparticle-catalyzed bimolecular reactions, $k_{\rm surf}$ may be used as free parameter \cite{roa2017} with Collins-Kimball boundary conditions as in unimolecular reaction.~\cite{CK}   

Let us recall that the limiting case $\ksurf/\kd = \infty$ in unimolecular reactions is typically called the `diffusion-controlled limit', where the reaction is limited only by the 3D diffusion of the reactants from the bulk space to the nanoparticle surface. In our case of bimolecular reactions, the 2D diffusion of the adsorbed reactants on the surface can be also limiting to the reaction as we will discuss in detail when we compare the hit-and-search to the hit-and-stick cases in the Results section.

\section{Results and Discussion}

\subsection{Total reaction rates}

\begin{figure*}[t]
 \includegraphics[width = 0.95\textwidth]{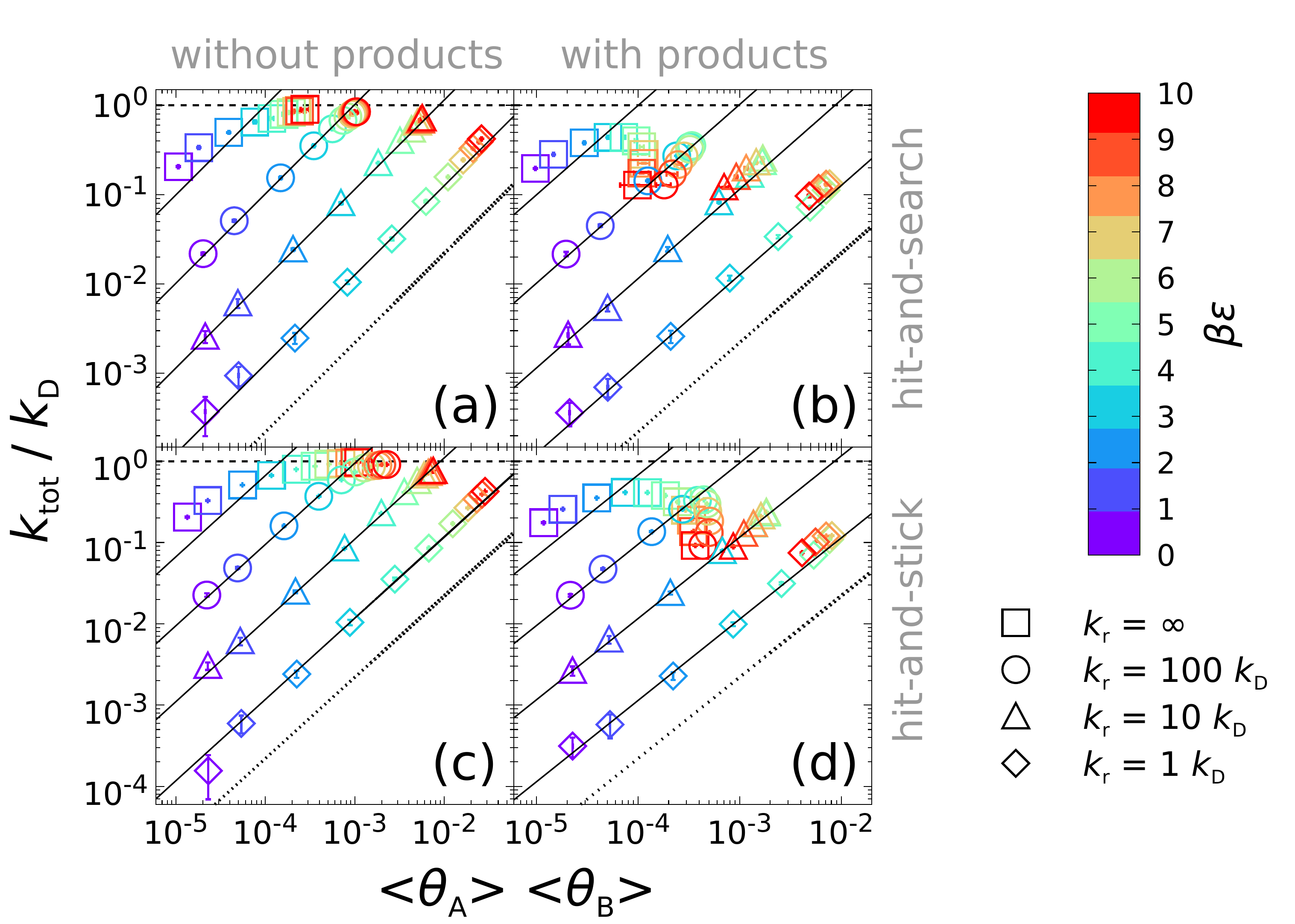}
 \caption{Total reaction rate of nanoparticle-catalyzed bimolecular reactions, $\ktot /\kd$, scaled by the 3D diffusion-controlled limit $\kd$ (represented by the horizontal black dashed line) versus the product of the mean coverages  $\langle \theta_{{\text{A}}}\rangle\langle \theta_{{\text{B}}}\rangle$. (a) Hit-and-search scenario without products. The black dotted line (bottom) is the reaction-controlled limit, eq.~(7). (b) Hit-and-search scenario with products. (c) Hit-and-stick scenario without products. (d) Hit-and-stick scenario with products. Solid lines are fits of the linear elementary law eq.~(2) to the data points for $\ep \leq 3~\kT$ (see Fig.~S2 in the Supporting Information). The different symbols depict different propensities $\ksurf$ and the color scheme illustrates the different adsorption strength $\varepsilon$ (see legends on the right hand side).}
 \label{fig:rate}
\end{figure*}

In Fig.~\ref{fig:rate}, we present the simulation results of the total reaction rate, $ k_{\rm tot}=({\rm d} n/{\rm d}t)/N_{\rm A}$, which measures the number of reaction events per unit time (see Fig.~S1 in the Supporting Information), normalized by the total number of reactant pairs, $N_{\rm A}=N_{\rm B}$, in our symmetric setup. The total rate $k_{\rm tot}$ is plotted as a function of the product of the adsorption coverages $\langle \theta_{{\text{A}}}\rangle\langle \theta_{{\text{B}}}\rangle$,  according to the standard elementary reaction law eq.~(2).  We scan through various microscopic reaction rates $\ksurf = 1, 10, 100, \infty$ in units of the unimolecular diffusion-controlled rate $\kd$ for unimolecular reactions, as well as different adsorption energies $\ep$ of the particles to the catalyst sphere. The rate $\kd$ is the upper limit of the total reaction rate in our model, that is, describing the rate of a particle that immediately reacts if it touches the catalyst surface. This fastest limit should be observable in the regime of high coverages at infinite propensity $\ksurf = \infty$, as we will in more detail discuss below. 

The scaled total rate $\ktot/\kd$ per particle is shown in Fig.~\ref{fig:rate}(a) versus $\langle \theta_{{\text{A}}}\rangle\langle \theta_{{\text{B}}}\rangle$ for the hit-and-search scenario (i.e., including surface diffusion) without products.  The adsorption energies are depicted with the color-code, and the symbols (see legends) show different values of the propensity $\ksurf$.  Note the log-log representation: the solid lines all with slope of unity reflect the linear scaling of the elementary reaction relation eq.~(2) with coverage, where we use the prefactor $k_\text{surf}$ as a fitting parameter (for details of fitting see Fig.~S2 in the Supporting Information). We see reassuringly that most of the data follow the linear scaling very well.  For very small propensities, $\ksurf = \kd$, i.e., closer to the reaction-controlled limit, we can try to quantify the apparent surface rate constant analytically: for slow reactions particles distribute as in equilibrium and the total rate can simply be estimated by the probability of two adsorbed reactants being in their reactive region, according to 
\begin{eqnarray}
k_{\rm tot} \simeq k_{\rm r} \frac{\pi(r_{\rm reac}-\sigma)^2}{N_{\rm A} 4\pi R_{\rm cat,eff}^2}\langle N_{\rm A_{\rm ad}}\rangle \langle N_{\rm B_{\rm ad}} \rangle =: k_{\rm surf}\langle \theta_{\rm A}\rangle \langle \theta_{\rm B}\rangle,
\end{eqnarray}
 essentially describing the ratio of accessible reactive area (corrected by excluded LJ size $\sigma$) to the total catalyst area. The apparent surface rate per particle can in this limit be expressed as $k_{\rm surf} = 64~k_{\rm r}(r_{\rm reac}-\sigma)^2R_{\rm cat,eff}^2/(N_{\rm A}\sigma^4)$.
  This prediction is shown in Fig.~\ref{fig:rate}(a) for $k_{\rm r} = k_{\rm D}$ with a dotted line and is indeed below and reasonably close to the simulation data.  Note for the lowest propensity ($\ksurf = \kd$), however, that for the highest coverages the behavior is slightly superlinear.   
  
  Further inspecting Fig.~\ref{fig:rate}(a), we see, as expected, that the overall reaction rate is increasing with larger the intrinsic propensity $\ksurf$.  The rate indeed reaches the fastest diffusion-controlled limit (1 in this normalized scale) for the larger $\ksurf$ and very large adsorptions. 
However, in the  diffusion-controlled limit, $\ksurf = \infty$, the functional rate behavior deviates qualitatively from the linear elementary law eq.~(2) for the large adsorption energies $\varepsilon \gtrsim 5~ \kT$, where $\ktot$ increases in a strikingly sublinear fashion with the coverages. Here, $\ktot$ saturates to the upper bound $\kd$ for $k_r=\infty$ and large coverages for $\varepsilon\gg ~\kT$. 
 
In Fig.~\ref{fig:rate}(b) we show $\ktot /\kd$ now including the products. The nonlinearities of the total reaction rate amplify and become very strong in the presence of the products. Products can stay adsorbed for a while for intermediate and high values of $\varepsilon$, which imposes additional correlations and a steric hindrance for the reactants on the nanoparticle during the reaction steady-state. (In enzyme kinetics such a steric hindrance leads to a product inhibition of the reaction and is typically modelled by additionally occupied binding sites in a Langmuir isotherm, which effectively decrease the reactant coverages.~\cite{enzyme}) The resulting states are highly non-trivial because the product surface coverages are self-consistently given by the rate of transformation of adsorbed reactants. Compared to Fig.~\ref{fig:rate}(a), less maximal coverages are reached, as could be intuitively expected due to steric (excluded volume) effects.  But note that simple 'mean-field' steric effects (like in a Langmuir isotherm) are already scaled away in our presentation by plotting the rate versus the explicitly realized coverages as output from the simulations.  In other words, if the products would simply rescale the coverage by excluded volume and not influence the prefactor $k_{\rm surf}$ in the eq.~(2), no change of the rates should be visible for fixed coverage. Hence, all differences between panels (a) and (b) are product effects beyond simple mean-field steric corrections.  

Remarkably, in the presence of products the nonlinearities of $\ktot$ are dramatically enhanced at large $\varepsilon$:  For the more diffusion-controlled reactions the total rate decreases sublinearly and even 'turns around' as a function of $\varepsilon$, that is, the rate has a maximum and both rate and coverage decrease for growing adsorption strength $\varepsilon$.  In contrast, for the more reaction-influenced reactions, where now also substantial non-linearities are present, the rates 'bend upward', while also reversing with increasing $\varepsilon$, again with the consequence of a maximum coverage along $\varepsilon$.  Hence, we find in general a maximization of reactant coverage and total rate versus adsorption energy in the region $\varepsilon \sim 5-7~\kT$. As an interesting consequence there is a bifurcation with respect to the coverage: {\it two different total rates can be established at the same reactant surface coverage} (see Fig. S6 in the Supporting Information for a close-up of the bifurcation).

\begin{figure}[t!]
 \centering
 \includegraphics[width = 0.45\textwidth]{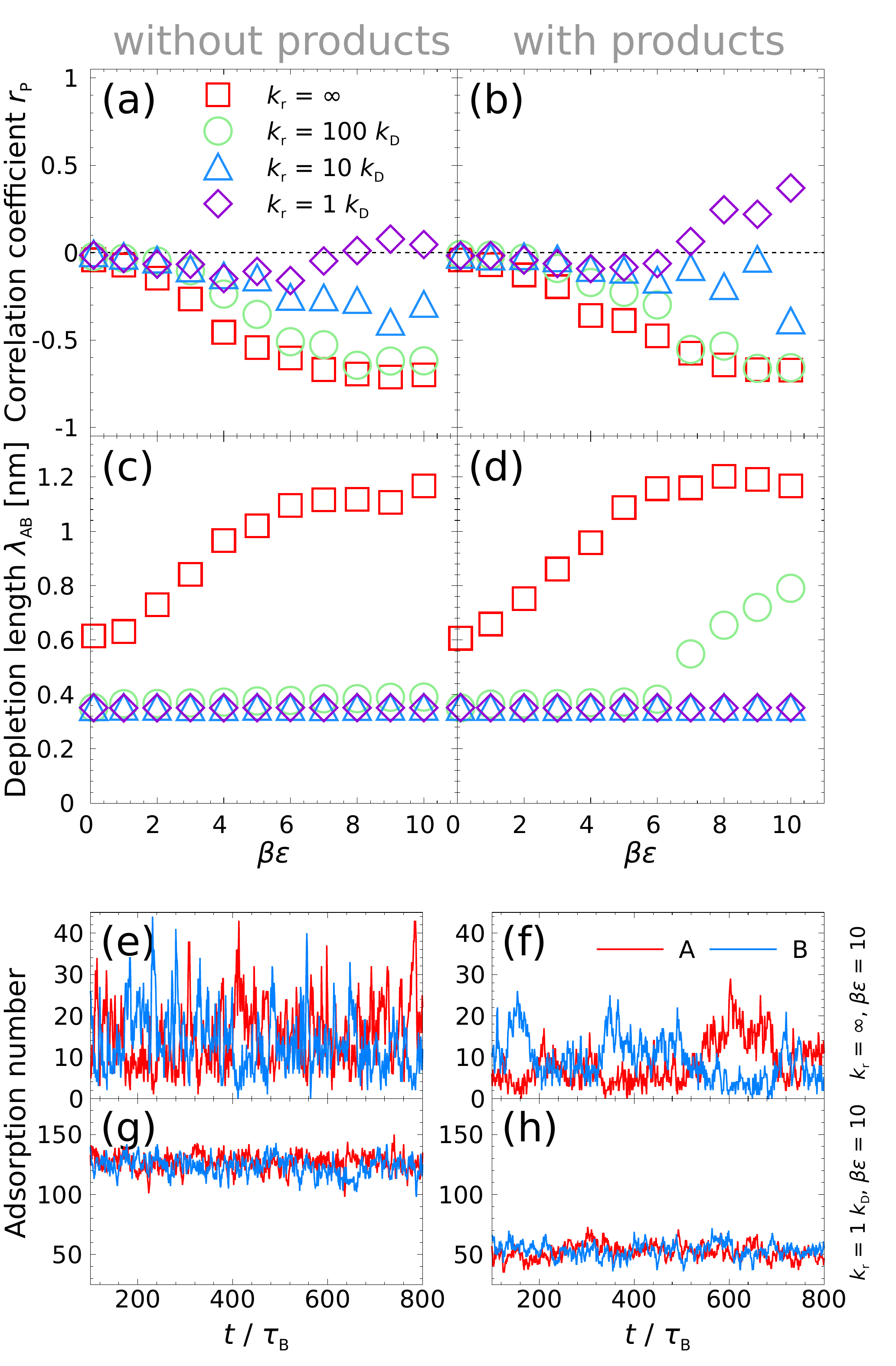}
 \caption{Reactant correlations on the nanoparticle surface in the hit-and-search scenario. (a) and (b): Number fluctuation (Pearson's) correlation coefficient $\rp$  of $N_{\rm A_{\rm ad}}$ and $N_{\rm B_{\rm ad}}$ versus  $\varepsilon$ for different chemical propensities $\ksurf$ without the products and with the products, respectively. (c) and (d):  Depletion length $\lab$ vs. $\ep$   without and with the products, respectively.  $\lab$ is calculated from surface RDFs  between A$_{\rm ad}$ and B$_{\rm ad}$ along the arc length on the spherical catalyst surface (see the Supporting Information). (e) and (f): The time series of adsorption numbers $N_{\rm A_{\rm ad}}(t)$ and $N_{\rm B_{\rm ad}}(t)$ for $\ksurf=\infty$ and $\varepsilon=10$ without and with products, respectively. (g) and (h): Time series as above but for $\ksurf=1~\kd$.}
\label{fig:corrwD} 
\end{figure}

In Figs.~\ref{fig:rate}(c) (without products) and (d) (with products) we present results on the hit-and-stick scenario (i.e., no surface diffusion), contrasting the free diffusion cases in (a) and (b).  The total rates are not influenced much by the immobile adsorbed reactants for the slow reactions, $\ksurf \leq 10\ \kd$;  here the process is mostly dominated by the microscopic reaction time and the 2D search does not play a big role.  However, the nonlinearities occurring for large adsorption energies $\varepsilon$ in the fast, more diffusion-controlled cases are more expressed than for the free surface diffusion.  Here, higher coverages are reached when compared to the mobile cases.  This we rationalize by the fact that the absence of surface mobility hampers the search for a reaction partner on the surface which is now a limiting factor for lower coverages. Thus, the reduced rates lead then to the higher coverage in the steady-state which than can re-establish the high rate again.  In Fig.~\ref{fig:rate}(d), we find compared to (c) that the effects of products are similar than for the case of including surface diffusion: rates are lower and nonlinearities more expressed, again leading to bifurcation phenomena. 

Noteworthy, in all cases of Figs.~2 (a)-(d) the unimolecular diffusion-controlled limit $\kd$ is not reached for infinite propensity $\ksurf = \infty$ for the low adsorption affinities. The reason is that reactants of one species still have to find another species on the catalyst surface, either by a 2D and/or 3D search on the surface~\cite{sano,freeman} or by a lucky hit during first approach. This scenario is reminiscent of the well-studied problem of a particle reacting with small reactive discs (in our case the adsorbed reaction partners), fixed and randomly distributed on an otherwise inert spherical sink (here, the nanoparticle), an important problem for ligands binding to receptors on the surface of biological cells.~\cite{Zwanzig,agarwal, Berez} In such a scenario, the diffusion-controlled reaction rate is about a factor $\simeq N_{\rm ad}r_{\rm reac} / (\pi R^{\rm eff}_{\rm cat})$ smaller than $k_{\rm D}$  in the limit of low coverage,~\cite{Zwanzig} where $N_{\rm ad}$ is the number of reactive sites of radius $r_{\rm reac}$. For our diffusion-controlled reactions in the low adsorption regimes, $N_{\rm ad}$ is on the order of 10 (see the Supporting Information), and with that the resulting rate indeed a factor 2-3 lower than the fasted limit $\kd$. Hence, $\ksurf\rightarrow \infty$ does not establish the unimolecular diffusion-controlled limit for low $\theta_i$ because of the diffusive search processes on the nanoparticle surface, which can be limited to low coverages. 


We will demonstrate in the following that the complex behavior of $\ktot$ can be attributed to a rich scenario of correlations and a resulting collective dynamical behavior of reactants (and products) on the surface of the catalyst.  This is to some extent already indicated in the snapshots in Fig.~\ref{fig:system}, which display spatial correlations in terms of domain formation or some extent of structuring of reactants,  particularly when the reactant and/or product coverages are large.  

\subsection{Spatial and dynamic correlations - heterogeneities in space and time}

To understand better the remarkably nonlinear behavior of the rates observed above we investigate correlations and fluctuations between the adsorbed reactants.  We first probe how the total numbers of adsorbed A and B particles on the surface are correlated by computing the fluctuation cross-correlation\cite{kuzovkov1988} (sometimes also  Pearson's~\cite{pearson}) coefficient $\rp (\varepsilon, \ksurf)= \langle (N_{\rm A_{\rm ad}} - \langle N_{\rm A_{\rm ad}}\rangle)(N_{\rm B_{\rm ad}}-\langle N_{\rm B_{\rm ad}}\rangle)\rangle/C$  where $C$ is a normalization factor to obtain numbers $-1 \leq \rp \leq 1$. This coefficient characterizes the two-point cross-correlations in the overall fluctuations of particle coverages with respect to the fully uncorrelated case. If the instantaneous excess coverage (over mean) of one species on the catalyst surface correlates favorably with the excess coverage of the other, then $\rp>0$. If $\rp<0$ they are anti-correlated. No correlations yields $\rp= 0$, i.e., excess coverages are fluctuating completely uncorrelated.  In Fig.~\ref{fig:corrwD} we show $\rp (\varepsilon)$ for different propensities $\ksurf$ for the hit-and-search scenario without (panel a)  and with (panel b) the presence of the products.   For the lowest chemical propensity ($\ksurf = \kd)$, hardly any correlations are visible for the case without products, while including products leads to positive correlations at high coverages.  For the more diffusion-controlled cases, $\rp$ decreases to large negative values, signifying that the adsorption fluctuations are strongly anti-correlated. These correlations at large adsorptions must be clearly related to the non-linearities of the reaction rates $\ktot$ in Fig.~\ref{fig:rate}. 


\begin{figure}[t!]
\centering
 \includegraphics[width = 0.48\textwidth]{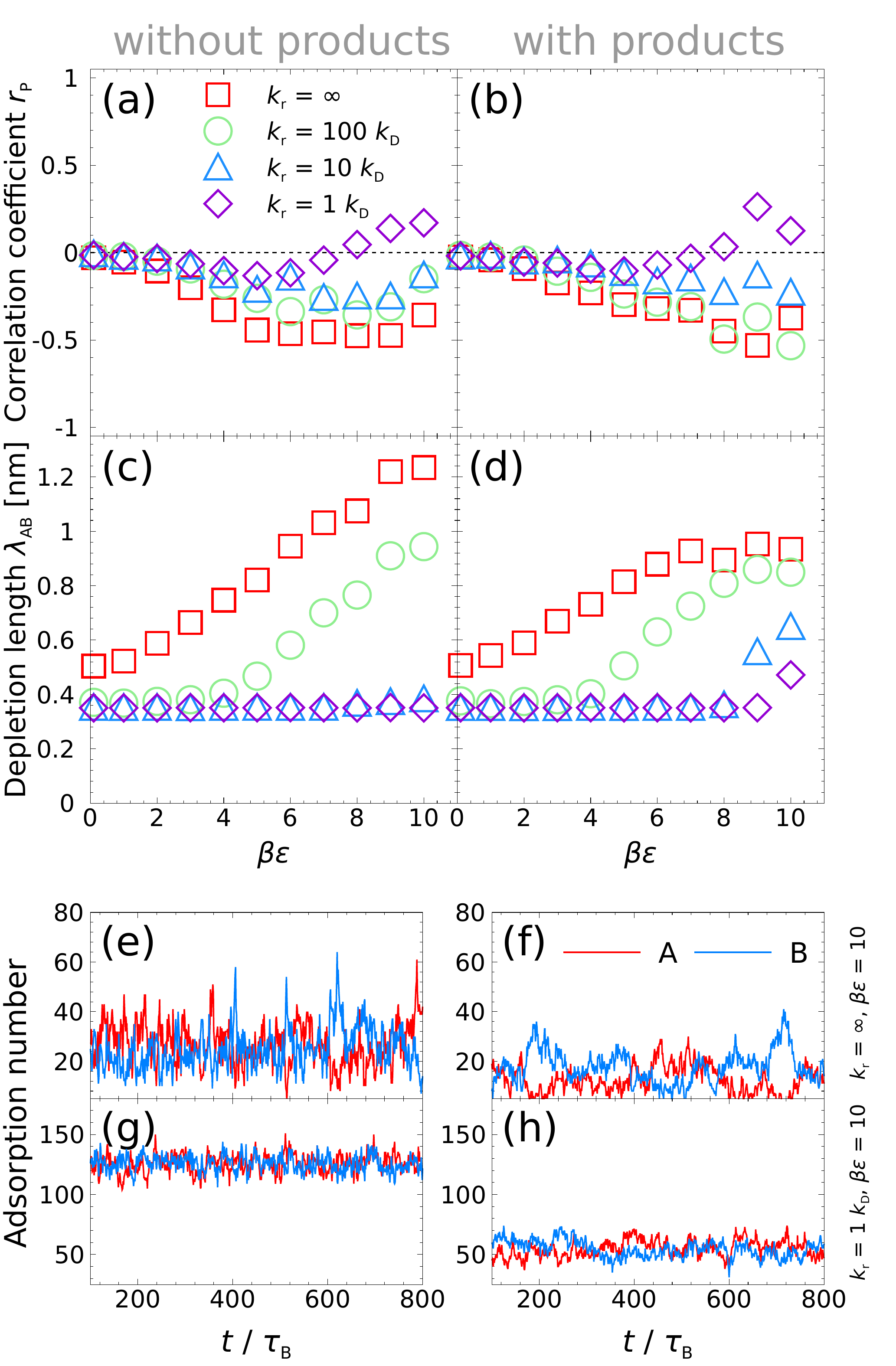}
 \caption{Same as Fig.~3 but now for the hit-and-stick scenario (i.e., without surface diffusion).}
 \label{fig:corrwoD}
\end{figure}

For a better interpretation we investigate how the number correlations are related to the local spatial particle arrangements of the reactants on the catalyst surface. For this, we define a depletion length $\lab (\varepsilon, \ksurf)$, which we estimate from the size of the correlation hole in the A$_{\rm ad}$-B$_{\rm ad}$ radial distribution functions (RDFs) along the catalyst arc length (being the same, in principle, as the decay length of spatial correlations of particles of the same kind, see Fig.~S3 in the Supporting Information for details). In Fig.~\ref{fig:corrwD} we show the depletion length $\lab (\varepsilon)$ for different propensities $\ksurf$ in the absence (panel c) and in the presence (panel d) of the products.  We find that for the low chemical reaction propensities the length $\lab$ stays around a constant value which is close to the particle size, indicating good mixing of A and B on the surface and close to mean-field behavior. For the slowest rates, the Pearson correlation is positive at high coverage and we observe large first peaks in the A-B RDFs (see Fig.~S4 in the Supporting Information), cf. snapshot Figs.\ref{fig:system}(b) (iii) and (iv).  The enhanced A-B spatial correlations (indicating better than ideal gas A-B mixing) must be made responsible for the slight, but observable super-linear rate behavior for the slowest reactions at high adsorption in the cases without products. In the presence of products the short-range correlations are amplified, probably due to the higher steric constraints and excluded-volume effects. 

However, when the reaction moves to the diffusion-controlled limit,  the depletion length $\lab$ increases substantially, especially if the adsorption $\varepsilon$ is strong, implying a growing domain formation and, considering the very negative Pearson coefficient, a kind of spatial micro-phase separation between the reactants A and B. In fact, closer inspection of the time series, representative examples shown in Figs.~3(e) and (f), indeed indicates multiple transient transitions towards a large depletion, even a full extinction of one species on the surface correlated to massive enrichment of the other, alternating quickly.  Products seem to quench these fluctuations, cf. Fig.~3(f), apparently stabilizing the demixed states for longer times. The anti-correlations are in stark contrast to the slightly positively correlated reaction-controlled cases as exemplified by the time series in Figs.~3(g) and (h)
(see more time series in Section VII and VIII of the Supporting Information for more details.)
 The different coverage correlations can also be well illustrated in a 2D density plot of $N_{\rm B_{\rm ad}}(t)$ versus $N_{\rm A_{\rm ad}}(t)$, cf. Fig.~S5 in the SI. 

\begin{figure*}[t!]
 \includegraphics[width = 0.8\textwidth]{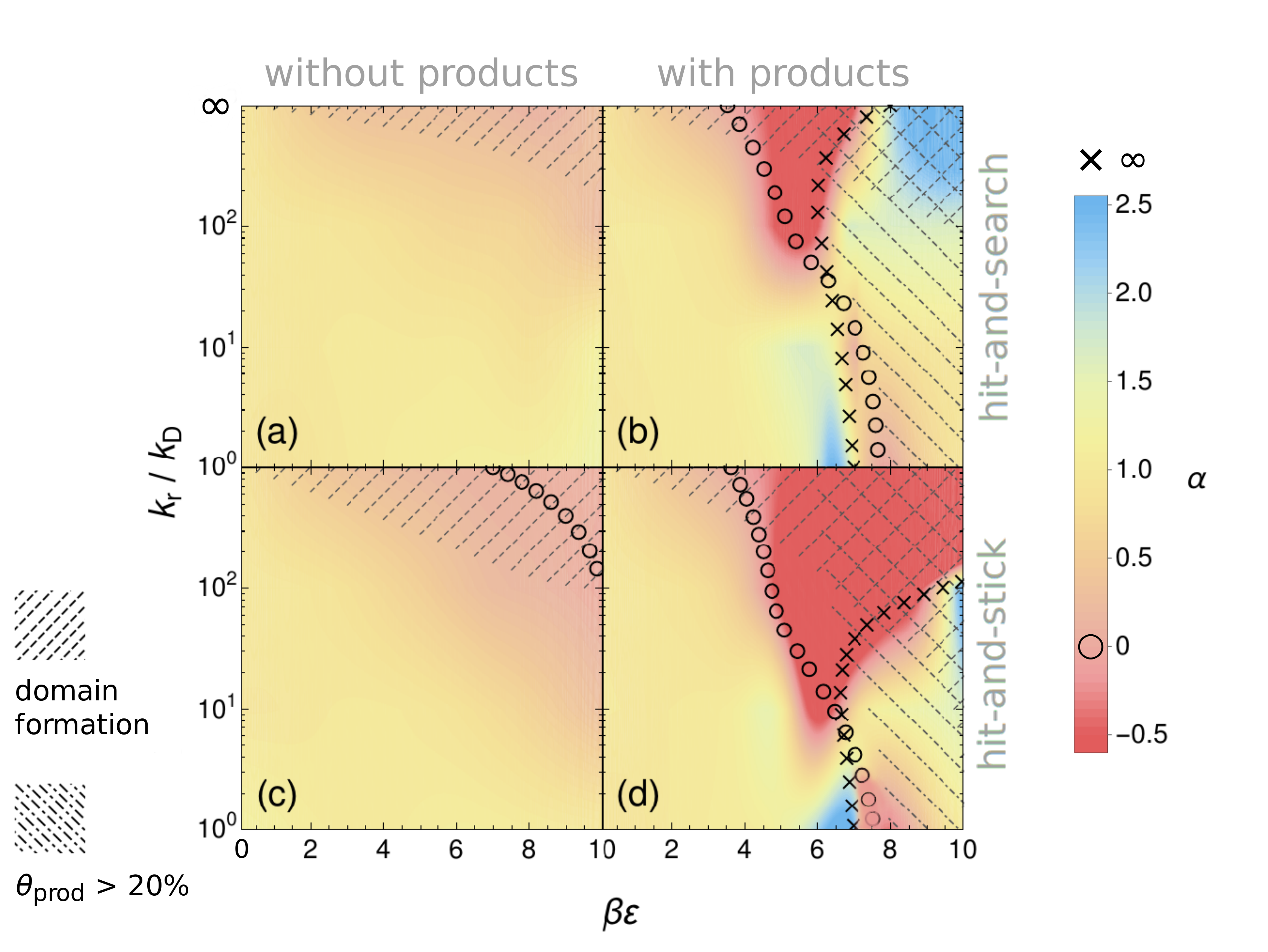}
 \caption{Apparent surface reaction rate order and dynamical state diagram. The colors encode the apparent order $\alpha = {\rm d}\log k_{\rm tot} /{\rm d}\log (\theta_{\rm A}\theta_{\rm B})$ of the reaction, see legend, plotted as a function of reaction propensity $k_{\rm r}/\kd$ and adsorption strength $\varepsilon$. Yellow is linear order $\alpha \simeq 1$, while the blueish shades depict super-linear behavior, $\alpha \gtrsim 1$, and the orange shades are sublinear behavior, $\alpha \lesssim 1$. Red shades depict a negative order, $\alpha < 0$.  The zeroth order line $\alpha = 0$ is signified by circle symbols ($\circ$), diverging order, $\alpha=\infty$, by the crosses ($\times$). The texture by dashed diagonal lines (at large propensities) mark regions of large domain deformation on the nanoparticle surface. The texture by dot-dashed diagonal lines (at large adsorption strength) mark regions of high product coverage ($\theta_{\rm prod} = \theta_{\rm C{\rm ad}}+\theta_{\rm D_{\rm ad}}>$~20\%). The plot is generated by interpolating values derived from  eq.~(8) based on all simulated data (see the Supporting Information).}
 \label{fig:alpha}
\end{figure*}

We performed the same analysis for the hit-and-stick scenario, cf. Fig.~\ref{fig:corrwoD}. The picture changes only quantitatively. For the more diffusion-controlled reactions the number fluctuations are still negatively correlated, albeit less expressed. The domain formation spatial range is enhanced in the immobile case, now even appearing also for slower reactions. We attribute the latter observation to the fact that without surface diffusion there is a lesser degree of mixing on the surface and the spatial demixing correlations can have a larger range. Importantly, the explicit 2D surface diffusion (as implemented in our model, still allowing a 3D hopping search) is apparently not critical for the observed effects and correlations.

\subsection{Brief discussion on transient domain formation}

The domain formation in the more diffusion-controlled systems can be observed nicely in the snapshots in Figs.~\ref{fig:system}(b) (i) and (ii).  The demixing is reminiscient of 'self-poisoning' by reactants in bistable surface reactions,\cite{Zhdanov1994} but here it is very transient due to the large nano-scale fluctuations, as indicated experimentally~\cite{nanofluct} and by approximate analytical theory~\cite{pineda}. Pattern formation and micro-demixing is a typical characteristic in reaction-diffusion systems and driven by local fluctuations, ~\cite{cross1993} similar found also in two-species Lotka-Volterra models.~\cite{tauber} The depletion and domain formation is obviously the source for the observed sublinear behavior of the total rate in Fig.~\ref{fig:rate}: The adsorbed reactants can only react at the 1D-interface lines between the domains, while the total coverage is fixed on average, resulting in a drastic change of reaction topology and dimensionality, sensitively depending on the coverage and rate themselves.  We note that we tried to characterize the stability of the domains but find only highly fluctuating, transient states for all parameters (see all time series in the Supporting Information). Inspecting a representative time series of the reaction rate (cf. Fig.~S1), we observe that rate fluctuations due to transient domain formations are very large and happen on the order of only a few $\tau_{\rm B}$ (the typical reactant diffusion time), that is, domain re-arrangements are indeed very fast and thus probably not easy to resolve in single-molecule experiments.~\cite{Buurmans,Peng:Nature:2008,Peng:PCCP:2009,Peng:JACS:2010} 
 
We note that the observed anti-correlated fluctuations and accompanying domain formations, that is, reactant segregation on the surface, is possible because of the high symmetry in the number of A and B particles in our systems. These effects are not observed in highly asymmetric A and B systems (numbers 50:450) for large adsorptions where the surface is mostly poisoned by the excess species B.

\subsection{Apparent surface reaction rate order and dynamical state diagram}

We now establish a surface reaction rate law behavior, in particular, to quantify the apparent order (scaling) of the total rate with coverages. The scaling behavior of the total rate in our symmetric (A$\equiv$B) system can be characterized by the general exponent $\alpha$ in the rate equation 
\begin{equation}
{\rm d}n/{\rm d}t = \kapp \left (\langle \theta_{{\text{A}}}\rangle\langle \theta_{{\text{B}}}\rangle\right)^\alpha, 
\end{equation}
that is, $\alpha$ represents the apparent order of the surface reaction reaction with respect to the produce of coverages. We have calculated $\alpha$ by evaluating the slopes of the data in Fig.~\ref{fig:rate}, i.e., $\alpha = {\rm d}\log k_{\rm tot} /{\rm d}\log (\theta_{\rm A}\theta_{\rm B})$. 

We are summarizing the behavior of the order $\alpha$ in the dynamical state diagram in Fig.~\ref{fig:alpha} by a color code plotted as a function of the scaled reaction propensity $\ksurf/\kd$ and adsorption strength $\beta \varepsilon$. The plot is generated by interpolating values derived from eq.~(8) based on all simulated data (see Figs.~S6 and S7 in the Supporting Information).
In the hit-and-search scenario without products displayed in panel (a), the picture as expected from conventional elementary reaction kinetics is mostly linear ($\alpha \simeq 1$; yellow), apart from the region of high adsorption and large propensities. The sublinear behavior (orange region; $\alpha \lesssim 1$), where the total rate and the coverages saturate, can be explained with the phenomenon of transient domain formation, which we depict by the textured region of diagonal dashed lines. The latter was estimated by average domain sizes $\lambda_{\rm AB} > r_{\rm reac}$.  In the hit-and-stick scenario without products, cf. Fig.~\ref{fig:alpha}(c), the picture changes only slightly, featuring a larger sublinear region and, in addition, a zeroth order line $\alpha = 0$, depicted by the circle symbols. The interpretation of the latter is that the rate does not change for an infinitesimal variation of the coverages (i.e., an infinitesimal change of the corresponding adsorption energy $\ep$).

The situation becomes far more complex when products are included, as shown in Figs.~\ref{fig:rate}(b) and (d). We empirically identify regions of large product feedback if the product coverage is more than 20\% of the catalyst surface area (see the Supporting Information). The latter regime is depicted by the textured parts with the diagonal dot-dashed lines and is being observed roughly for $\varepsilon \gtrsim 5~\kT$. As discussed above, we see a turnover behavior and bifurcation of $\ktot$, leading to singularities where the apparent reaction order can even diverge, $\alpha = \infty$, depicted by cross symbols in the state diagram. We find in Figs.~\ref{fig:rate}(b) and (d) that all orders are essentially observed. While zeroth and diverging orders are close by for adsorptions between 4 and 7 $\kT$, they change their order in appearance versus $\varepsilon$ depending on the chemical propensity. The complexity of the diagram reflects the coupling of various spatial correlations and product inhibition on the surface. Note also that there is a re-entrant behavior with larger adsorption, i.e., for $\varepsilon \gtrsim 8$~$\kT$ the order becomes linear ($\alpha=1$) again.  The situation stays similarly complex for the hit-and-stick scenario, see panel (d). The system is now more affected by product inhibition, leading to a wider sublinear region.

As summarized in the state diagram, the rate law for bimolecular elementary reactions as in eq.~(2) is violated for larger adsorptions $\varepsilon \gtrsim 5~\kT$,  in particular for diffusion-influenced reactions where large fluctuations lead to transient domain formation. Including products that stay adsorbed on the nanoparticle surface renders the kinetics much more complex, going beyond simple steric packing effects. The presence of products can lead to a bifurcation effect (indicated by crosses in Figs.~\ref{fig:rate}(b) and (d)) where the same rate can be established by two different coverages, and increasing adsorption (decreasing temperature) can lead to decelerating rates (dark red regions Figs.~\ref{fig:rate}(b) and (d)). Quantitatively, all these effects may be model-dependent, but on a qualitative level they may survive in 'real' experimental nanoparticle catalysis and thus add additional complexity beyond more chemical and structural surface properties, e.g, defects, chemical functionalizations, reconstructions, etc.  

\section{Concluding Remarks}

We have studied bimolecular reactions catalyzed by a perfectly spherical model-nanoparticle using interacting particle-resolved reaction-diffusion computer simulations. We found a rich scenario of physical phenomena intrinsic to the reaction-diffusion process due to large fluctuations and steric correlations in the systems, which lead to a highly nonlinear behavior of the rates versus surface coverages. In particular, in rapidly reacting and crowded situations the nonlinear behavior is most strongly expressed and significantly modified by the the presence of reaction products.  

A direct comparison to experiments is very difficult. Even though from single molecule experiments~\cite{Buurmans,Peng:Nature:2008,Peng:PCCP:2009,Peng:JACS:2010} direct information on the heterogeneity in time and space can be measured, detailed interpretations and resolution of all time scales are not always possible. Often it is argued that NP surface effects, such as reconstruction, are the reason for the large fluctuations and heterogeneities. We emphasize again that the observed complex phenomena in our catalyst model with a perfectly smooth and inert nanoparticle surface are all based on reaction-diffusion nano-scale fluctuations and not due to specific surface features (such as facets, defects, reconstruction, etc.), as often reported in literature. In other words, in some time and space regimes fluctuations and heterogeneities are intrinsic to the reaction-diffusion problem, and not due to any surface effects. Hence, our results call for a careful re-interpretation of some existing experimental data in nanoparticle catalysis as well as for conducting new experiments which could be designed to systematically probe and resolve the observed effects.

We also hope that the results inspire further development and extensions of related nonlinear rate theory, e.g., using recent promising first-passage time approaches,~\cite{benichou2005,benichou2010,benichou2011,grebenkov} as well as additional particle-resolved simulations employing, for example, more realistically resolved reactants, products, and nanoparticles.  In particular, the action of adsorbed and crowding products is remarkably non-trivial and in general goes beyond simple steric (Langmuir) packing effects. 

In our model the reactant coverage is the decisive parameter tuned by adsorption energies of the reactants to the catalyst surface, which may be related to effects of temperature changes visible in experiments in more diffusion-influenced reactions. We expect all the nonlinear phenomena also to appear for varying bulk concentrations, which shall be tested systematically in future computer simulations or experiments. 

\normalsize
\subsection*{Conflict of Interest} The authors declare no competing
financial interest.

\subsection*{Supporting Information Description}
Details on the calculation of rates, scalings, and the radial distribution functions. Mean adsorptions and coverages of reactants and products and their variances. Selected trajectory time series for coverages of reactants and products.  

\subsection*{Acknowledgments}
 The authors thank Yonghyun Song and Matthias Ballauff for inspiring discussions, and Matej Kandu{\v c}, Rafael Roa, and Stefano Angioletti-Uberti for a critical reading of the manuscript. This project has received funding from the European Research Council (ERC) under the European Union's Horizon 2020 research and innovation programme (grant agreement Nr. 646659). W.K.K. acknowledges the support by a KIAS Individual Grant (CG076001) at Korea Institute for Advanced Study.

\footnotesize
\setlength{\bibsep}{0pt}

\bibliography{nanoparticle_paper_YC_preprint}

\clearpage

\makeatletter
\setlength\acs@tocentry@height{8.25cm}
\setlength\acs@tocentry@width{4.45cm}
\makeatother

\section*{TOC Graphic}
\centering
\includegraphics[width = 8.1cm]{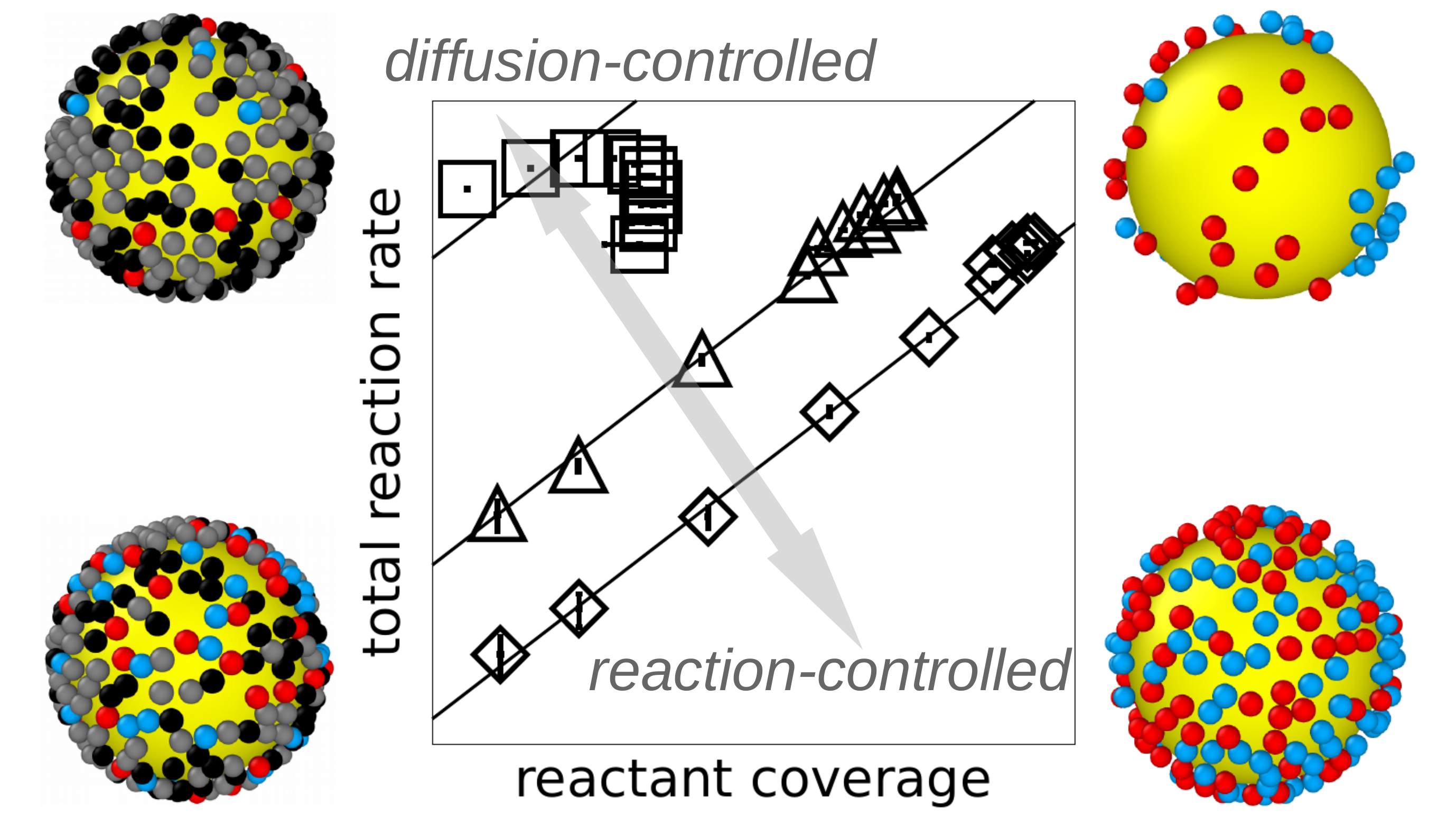}

\end{document}